\documentclass[sigconf]{acmart}

\AtBeginDocument{%
  }

\copyrightyear{2026}
\acmYear{2026}
\setcopyright{cc}
\setcctype{by}
\acmConference[LAK 2026]{LAK26: 16th International Learning Analytics and Knowledge Conference}{April 27-May 01, 2026}{Bergen, Norway}
\acmBooktitle{LAK26: 16th International Learning Analytics and Knowledge Conference (LAK 2026), April 27-May 01, 2026, Bergen, Norway}
\acmPrice{}
\acmDOI{10.1145/3785022.3785049}
\acmISBN{979-8-4007-2066-6/2026/04}

\begin{document}

\title{Modeling Collaborative Problem Solving Dynamics from Group Discourse: A Text-Mining Approach with Synergy Degree Model}


\author{Jianjun Xiao}
\orcid{0000-0003-0000-9630}
\affiliation{%
  \institution{Beijing Normal University}
  \city{Beijing}
  \country{China}}
\email{et\_shaw@126.com}

\author{Cixiao Wang}
\authornote{Corresponding author.}
\orcid{0000-0002-6053-8991}
\affiliation{%
  \institution{Beijing Normal University}
  \city{Beijing}
  \country{China}}
\email{wangcixiao@bnu.edu.cn}

\author{Wenmei Zhang}
\orcid{0009-0003-0707-8898}
\affiliation{%
  \institution{South China Normal University}
  \city{Guangdong}
  \country{China}}
\email{zwmwenmei@163.com}

\renewcommand{\shorttitle}{Synergy Degree Model of Collaborative Problem Solving in cMOOCs}

\begin{abstract}
Measuring collaborative problem solving (CPS) synergy remains challenging in learning analytics, as classical manual coding cannot capture emergent system-level dynamics. This study introduces a computational framework that integrates automated discourse analysis with the Synergy Degree Model (SDM) to quantify CPS synergy from group communication. Data were collected from 52 learners in 12 groups during a 5-week connectivist MOOC (cMOOC) activity. Nine classification models were applied to automatically identify ten CPS behaviors across four interaction levels: operation, wayfinding, sense-making, and creation. While BERT achieved the highest accuracy, GPT models demonstrated superior precision suitable for human–AI collaborative coding. Within the SDM framework, each interaction level was treated as a subsystem to compute group-level order parameters and derive synergy degrees. Permutation tests showed automated measures preserve construct validity, despite systematic biases at the subsystem level. Statistical analyses revealed significant task-type differences: survey study groups exhibited higher creation-order than mode study groups, suggesting “controlled disorder” may benefit complex problem solving. Importantly, synergy degree distinguished collaborative quality, ranging from excellent to failing groups. Findings establish synergy degree as a sensitive indicator of collaboration and demonstrate the feasibility of scaling fine-grained CPS analytics through AI-in-the-loop approaches.
\end{abstract}




\maketitle

\section{Introduction}

Collaborative problem solving (CPS) is widely recognized as a critical twenty-first century competence \citep{oecd_pisa_2017, fiore_collaborative_2018}. With the proliferation of digital discourse, CPS provides rich opportunities for fine-grained interaction analysis in learning analytics. A central challenge, however, lies in modeling collaboration dynamics beyond simple frequency counts or static measures. Classical approaches, such as manual coding based on the Interaction Analysis Model (IAM) \citep{gunawardena_analysis_1997} or the Connectivist Interaction and Engagement framework (CIE) \citep{wang_interaction_2017}, are not only time-consuming but also insufficient for capturing the systemic and emergent nature of collaboration \citep{tian_measurement_2025}. Given that CPS is inherently a complex system where behaviors interact and self-organize into performance-driving patterns \citep{ouyang_artificial_2023}, computational models are needed to represent CPS as integrated systems and to quantify synergy effects.  

Recent advances in natural language processing (NLP) have enabled automated identification of CPS behaviors from collaborative discourse \citep{wang_role_2025}. Large language models (LLMs) such as BERT and GPT can classify utterances with high accuracy, substantially reducing the reliance on manual annotation \citep{yan_practical_2024}. At the same time, the Synergy Degree Model (SDM) offers a means to measure both the order of individual behavioral subsystems and the overall synergy effect arising from their interactions \citep{haken_synergetics_1977, kong_examining_2025}. Applying SDM to CPS conceptualizes behavior types as interacting subsystems, allowing the calculation of order parameters and the estimation of system-level synergy.  

However, empirical applications of SDM in CPS remain limited. Existing studies tend to emphasize descriptive analyses \citep{chai_computer-based_2023} or network structures \citep{zhang_understanding_2022, feng_group_2025}, rather than capturing systemic synergy. Exploring how synergy patterns vary by problem type (such as task complexity and structuredness) and performance level could provide actionable insights for designing effective collaborative learning environments \citep{ouyang_artificial_2023, almaatouq_task_2021}.  

To address these gaps, this study proposes a text-mining approach to automatically identify CPS behaviors and quantify their dynamics using SDM. We address three research questions:  

\begin{itemize}
    \item \textbf{RQ1:} How can CPS key behaviors be automatically identified from collaborative discourse?  
    \item \textbf{RQ2:} How can the synergy effect of key CPS behaviors be quantified using SDM?  
    \item \textbf{RQ3:} Do groups with different problem types and collaborative quality levels exhibit significant differences in order parameters and synergy degree in the CPS-orientated SDM? 
\end{itemize}  

This study makes three main contributions. First, it integrates NLP-based behavior coding with system-level synergy modeling. Second, it introduces order parameters and synergy degree as indicators of CPS behavioral coordination. Third, it provides empirical evidence linking problem type and collaborative quality to CPS synergy. Together, these contributions demonstrate how text-based learning analytics can be leveraged to model complex group collaboration dynamics.
\section{Related Work}

\subsection{Conceptualizing CPS as an Open System in cMOOC}
CPS is commonly described as a process in which multiple individuals, operating under shared goals and conditions of interdependence, coordinate social interaction and cognitive reasoning to represent the problem, generate and evaluate solution options, and arrive at a resolution \citep{omalley_construction_1995}. The concept comprises two core facets: cognitive mechanisms involving joint problem representation, evidence integration, and metacognitive monitoring; and social mechanisms involving role distribution, conflict negotiation, and joint attention \citep{barron_when_2003}. However, classical process-oriented perspectives—whether focusing on socio-cognitive processes \citep{rose_analyzing_2008}, regulatory mechanisms \citep{lazakidou_using_2010}, or interaction quality \citep{weinberger_framework_2006}—tend to treat CPS as a sequential procedure rather than an integrated system, struggling to capture emergent coordination that arises from interactions among multiple behavioral components \citep{jacobson_conceptualizing_2016}.

An open system perspective offers a complementary lens: CPS can be conceptualized as a complex adaptive system that continuously exchanges information with its environment, exhibits nonlinear dynamics, and generates emergent properties through self-organization \citep{haken_synergetics_1977, dron_pedagogical_2023}. This view aligns naturally with connectivism, which conceptualizes learning as the formation, maintenance, and reorganization of connections among nodes within distributed networks, emphasizing information fluidity and continual network optimization \citep{Siemens2005}. Based on the connectivist principle, learners in cMOOCs generate new ideas through self-organizing interaction with peers and technology, aggregating heterogeneous information and producing knowledge artifacts through processes of \textit{aggregation}, \textit{remixing}, \textit{repurposing}, and \textit{feed-forwarding} \citep{bakki_moocat_2019, downes_connectivism_2022, kop_connectivism_2008}. These characteristics—continuous information exchange, self-organization, and emergent knowledge creation—position cMOOC-based CPS as an open system par excellence.  

The CIE framework \citep{wang_interaction_2017} provides an operational basis for modeling CPS as a multi-subsystem structure. This framework identifies four interaction types according to cognitive engagement levels: \textit{operation} (engagement with technological environment), \textit{wayfinding} (establishing social connections and sharing information), \textit{sense-making} (constructing meaning through negotiation), and \textit{creation} (producing knowledge artifacts). From an open system perspective, these four interaction types can be conceptualized as interdependent subsystems whose coordinated evolution characterizes overall collaborative dynamics. The framework has been applied in cMOOC studies examining content production \citep{xu_research_2024, bai_impact_2021}, interaction patterns \citep{xu_research_2023}, and CPS \citep{zhang_comparative_2024}, using methods such as content analysis, SNA, and ENA \citep{tian_measurement_2025, guo_analysis_2025}. However, research on automatically identifying CPS behaviors and, more importantly, on measuring the synergistic effects among these behavioral subsystems remains limited. This gap is critical to capturing the system-level dynamics that drive collaborative performance.  

\subsection{Automated Coding of Collaborative Discourse}
Automated coding of collaborative discourse aims to identify and quantify the socio-cognitive functions within group interaction using reproducible computational methods, thereby providing actionable evidence for process-oriented assessment and adaptive support. Along the recent trajectory of technological advances, existing approaches can be broadly categorized into three classes—classical machine learning, deep learning, and LLMs \citep{rose_analyzing_2008, de_wever_content_2006, reimann_time_2009, lang_handbook_2022}. Among them, the application of LLM (e.g., BERT, GPT) by researchers has become the mainstream of educational tasks \citep{yan_practical_2024}.

Classical machine learning centers on feature engineering. It combines textual features with conversational structure features and employs classifiers such as support vector machines \citep{kovanovic_automated_2014} and random forests \citep{kovanovic_towards_2016} to perform turn-level classification or sequence labeling \citep{xing_beyond_2019, liu_quantifying_2021}. With the advent of deep learning, researchers have adopted techniques such as FastText, TextCNN, and TextRNN for text analysis, thereby enhancing the robustness of automated discourse coding. The development of LLMs has led to substantive advances in the scalability, adaptability of automated functional coding and structure extraction for collaborative discourse. On the one hand, some scholar proposed a new variant of the MOOC-BERT model for cognitive presence recognition based on large-scale unlabeled discussion data from numerous MOOCs involving various disciplines \citep{liu_mooc-bert_2023}, which is significantly better than classical machine learning models. Other scholars have developed BERT-CNN–based models for identifying emotional and cognitive engagement, incorporating explainable AI (xAI) techniques to interpret the results \citep{liu_automated_2022}. On the other hand, with the continuous improvement in the performance of GPT models, an increasing number of studies have begun to employ GPT under zero- or few-shot conditions to encode and analyze learners’ discourse \citep{wang_evaluating_2025, hu_exploring_2025}. Although GPT has been applied to text classification tasks, its complexity makes it difficult to effectively employ existing xAI methods, thereby limiting deeper understanding of its decision-making mechanisms. These studies provide a reference for the automatic classification of CPS key behaviors in cMOOC. Therefore, we will compare the classification performance of classical models, neural networks and LLMs.

\subsection{From Process Analytics to System Dynamics in CPS}
Research on learning analytics for CPS can be characterized as a dual-path evolution from “computational analytics” to “analytics for collaboration,” advancing along the trajectory of “static profiling—temporal processes—complex system organization” \citep{chen_learning_2020, lang_handbook_2022}. 

In the early stage, readily available behavioral and discourse data (e.g., number of posts and replies, social network centrality, discourse levels) were used to depict participation, and were linked via correlational or predictive models to test and assignment outcomes, thereby revealing an overall “activity–outcome” association \citep{macfadyen_mining_2010}. However, this line of work struggled to explain why collaboration advances, stalls, or fails at particular times—that is, it fell short of providing generative, mechanism-level explanations. Researchers have introduced xAI techniques to address these challenges. For example, some scholars have developed interpretable machine learning models based on multidimensional features of students’ collaborative processes to automatically identify learners’ collaborative roles \citep{wang_role_2025}. Subsequently, the focus shifted to temporal process modeling in authentic contexts such as classroom small groups, PBL, and online seminars, introducing methods such as sequence mining \citep{hoppe_using_2021}, Markov models \citep{saqr_when_2023}, and transition graphs \citep{saqr_transition_2025} to reveal and explain the relationship between "process" and "performance." Further, researchers have employed computational linguistics approaches to identify collaborative dynamics from participants’ discourse. For example, the Group Communication Analysis (GCA) framework has been proposed to detect the emergence of learners’ roles during collaboration \citep{dowell_group_2019}. Other scholars have applied the ENA method to evaluate the connections and structures of meaning elements within collaborative discourse \citep{zhang_understanding_2022}. Building on this, recent AI-driven frameworks combine multichannel sequences, clustering, linkage structures, and latent-state models to characterize the emergent organization of team collaboration and to establish interpretable connections with learning outcomes \citep{ouyang_artificial_2023}. 

Overall, the research paradigm has shifted from "outcome association" toward "process mechanism," and further toward a "complex adaptive systems" perspective: collaborative performance is not a linear function of any single dimension of activity, but rather emerges from the coordinated organization of multiple functional cues. Correspondingly, measurement has moved from "how many times" to "how the system is organized and when it switches," supplemented by multimodal indicators to explain nonlinear transitions \citep{saqr_how_2022}. This perspective provides actionable parameters and triggers for comparable diagnostics across tasks and contexts, as well as for micro-interventions at critical junctures, and aligns with design dimensions in "analytics for collaboration" regarding the role of analytics (partner/regulator) and the degree of coupling in the workflow (loose/tight) \citep{chen_learning_2020}. Nevertheless, there remain critical gaps in the operationalization and validation of system-level collaborative indicators, which constrain the translation from discourse evidence to system-level diagnosis and intervention. For practice-oriented CPS learning analytics, it is essential to construct an operational pipeline "from discourse to system dynamics" \citep{park_discourse_2025}: beginning with scalable automated discourse coding, identifying key collaborative behaviors from discourse, and establishing paradigms for robust evaluation, thereby enabling comparable diagnostics across tasks and levels of collaboration quality. Therefore, this study aims to explore how to model CPS dynamics from discourse data and to further analyze the relationships between these dynamics, task types, and collaborative performance.
\section{Method}

\subsection{Research Context and Participants}

This study was conducted in a Chinese cMOOC titled \textit{Internet+Education: Dialogue between Theory and Practice}, explicitly designed based on connectivist principles. The CPS activity spanned five weeks: Week 0—problem presentation and group formation; Week 1—problem analysis and planning; Week 2—resource connection and processing; Week 3—instructor facilitation and refinement; and Week 4—product submission and reflection. 12 authentic, ill-structured problems were designed for \textit{"Internet+Education"} and were classified into three categories: survey study, mode study, and design study. Specifically, a survey study refers to collecting data on a particular topic to analyze its current status and developmental trends; a mode study involves analyzing and synthesizing the concepts, connotations, essence, models, mechanisms, and underlying patterns of a given phenomenon; and a design study refers to designing, implementing, and evaluating a tool for a specific context or target group.   

A total of 52 learners participated and self-organised into 12 groups (G1–G12) with 3–7 members per group in an open online learning environment. As shown in Table~\ref{tab:group_composition}, groups varied in composition: approximately half homogeneous (similar backgrounds) and half heterogeneous (diverse backgrounds). Collaborative quality refers to each group’s overall performance—both in process and outcome—during the problem-solving stages. After the course concluded, the course team conducted a comprehensive evaluation of each problem-solving group’s collaborative quality based on their live presentations, process participation, completion of platform tasks, reflection reports, interim outputs, and final products. The groups were then rated at four levels: excellent, good, pass, and failed.     

\begin{table*}[htbp]
\centering
\small
\caption{Group composition, problem types, collaboration quality, and homogeneity in the CPS stage}
\label{tab:group_composition}
\begin{tabular}{c|c|c|c|c|c}
\toprule
\textbf{Group} & \textbf{Problem Type (PT)} & \textbf{Group Composition (GC)} & \textbf{Homogeneity} & \textbf{Collaboration Quality (CQ)} & \textbf{Number of Members} \\
\midrule
G1  & SS     & Industry-mixed         & Hetero. & Failed    & 3 \\
G2  & DS     & Teacher–Student Mixed  & Hetero. & Pass      & 5 \\
G3  & SS     & Industry-mixed         & Hetero. & Excellent & 3 \\
G4  & MS       & All Students           & Homo.   & Excellent & 4 \\
G5  & MS       & Teacher–Student Mixed  & Hetero. & Pass      & 5 \\
G6  & SS     & All Students           & Homo.   & Good      & 7 \\
G7  & SS     & All Teachers           & Homo.   & Good      & 3 \\
G8  & SS     & All Students           & Homo.   & Pass      & 3 \\
G9  & SS     & Teacher–Student Mixed  & Hetero. & Excellent & 6 \\
G10 & DS     & Industry-mixed         & Hetero. & Good      & 3 \\
G11 & SS     & Teacher–Student Mixed  & Hetero. & Good      & 5 \\
G12 & DS     & All Students           & Homo.   & Excellent & 5 \\
\bottomrule
\end{tabular}
\\[2mm]
\raggedright \footnotesize \textit{Note.} Homo. = Homogeneous; Hetero. = Heterogeneous group composition. Task complexity is determined by the problem type: survey study (SS) is considered simple, whereas mode study (MS) and design study (DS) are classified as complex.
\end{table*}

\subsection{Research Design and Procedure}

\begin{figure*}
    \centering
    \includegraphics[width=1\linewidth]{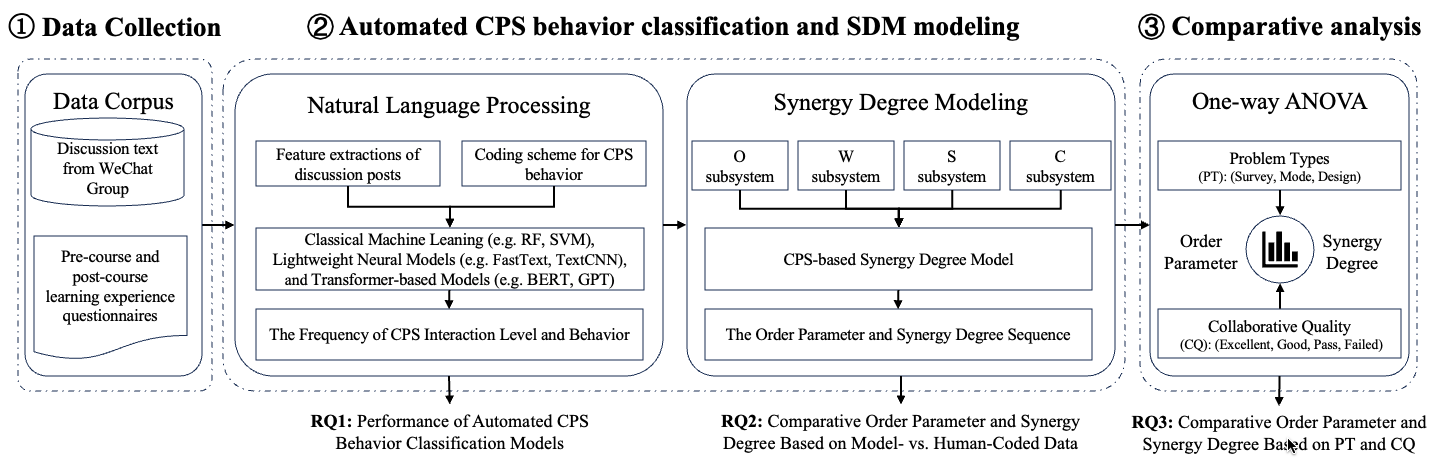}
    \caption{The research design and procedure}
    \label{fig:research_design}
\end{figure*}

The research design followed a three-phase process combining data collection, automated CPS behavior classification with SDM modeling, and comparative analysis (Fig.~\ref{fig:research_design}).

\textbf{Phase 1: Data collection and cleaning.}  
Data were collected from the cMOOC platform (\url{https://cmooc.bnu.edu.cn}) during the problem-solving stage, including group chat logs, discussion posts, and composition information. Data quality procedures included removing irrelevant messages, merging fragmented utterances, and anonymizing participant identifiers. The resulting dataset captured authentic collaborative discourse for all 12 groups.

\textbf{Phase 2: Automated CPS behavior classification and SDM modeling.}  
We adopted a coding framework covering four CPS interaction categories: operation, wayfinding, sense-making, and creation. Multiple models were compared: classical ML (SVM, Decision Tree, Random Forest, KNN), neural models (FastText, TextCNN, TextRNN), and LLMs (BERT, GPT). 5-fold cross-validation was employed given the sample size ($N = 2,420$). Performance was evaluated using weighted precision, recall, and F1-score due to class imbalance.

After obtaining behavior-labeled data, we constructed the CPS-orientated SDM (detail see Sec. \ref{sec:synergy_degree_model}) to compute (a) \textit{order parameters} of each CPS behavior category and (b) \textit{overall synergy degree} at group level. Permutation tests \citep{boda_short-term_2020} assessed statistical differences between human-annotated and model-predicted data.

\textbf{Phase 3: Comparative analysis.}  
Statistical analyses examined differences in CPS order parameters and synergy degree across \textit{problem types} (survey, mode, design) and \textit{collaborative quality} (excellent, good, pass, fail). Assumptions relevant to ANOVA – such as normality and homogeneity of variance – were evaluated before applying ANOVA or Kruskal-Wallis tests, followed by post-hoc comparisons when appropriate. 

\subsection{CPS behavior coding}

\subsubsection{Design of the CPS Coding Framework} \label{sec:coding_codebook}

The coding framework was developed based on connectivist learning principles and prior collaborative learning research \citep{wang_interaction_2017}, consisting of 10 CPS behavior categories organized into four task-related interaction levels plus irrelevant interaction (Table~\ref{tab:cps_coding_scheme}). Specifically, \textbf{operational interaction} refers to creating and maintaining shared collaborative space: \textit{space setup} (O1) and \textit{technical operation} (O2); \textbf{wayfinding interaction} captures building interpersonal connections and information sharing: \textit{social connection} (W1), \textit{content link} (W2), and \textit{task alignment} (W3); \textbf{sense-making interaction} reflects collective information processing and meaning negotiation: \textit{idea suggestion} (S1), \textit{conflict negotiation} (S2), and \textit{planning/decision} (S3); \textbf{creation interaction} refers to knowledge integration and shared artifact production: \textit{integrative creation} (C1); \textbf{irrelevant interaction} (I) captures task-unrelated messages.

\begin{table*}[htbp]
\centering
\small
\caption{Coding scheme for CPS behaviors with examples}
\label{tab:cps_coding_scheme}
\begin{tabular}{p{2.5cm}p{2.5cm}p{0.8cm}p{6cm}p{4cm}}
\toprule
\textbf{Interaction Level} & \textbf{CPS behavior} & \textbf{Code} & \textbf{Description} & \textbf{Example} \\
\midrule
Irrelevant Interaction (I) & Irrelevant Info & I & Operations or messages unrelated to the task & "Message was withdrawn" \\
\midrule
Operational Interaction (O) & Space Setup & O1 & Initiating chat groups, creating shared documents & "I created a group, please join" \\
                       & Technical Operation & O2 & Technical guidance on platform use, tools & "How to share the screen?" \\
\midrule
Wayfinding Interaction (W)  & Social Connection & W1 & Greetings, welcoming, checking availability & "Hi everyone!", "Are you free tomorrow?" \\
                       & Content Link & W2 & Sharing resources, providing/asking basic info & "Here is a related paper I found" \\
                       & Task Alignment & W3 & Reporting progress, encouraging peers & "I have finished my part, keep going!" \\
\midrule
Sense-making Interaction (S) & Idea Suggestion & S1 & Offering opinions, suggestions, recommendations & "We can try case analysis, it fits our problem" \\
                        & Conflict Negotiation & S2 & Raising questions, pointing out disagreements & "I disagree, let's discuss the reasons" \\
                        & Planning/Decision & S3 & Coordinating tasks, deciding methods & "Let's meet Wednesday to finalize the plan" \\
\midrule
Creation Interaction (C) & Integrative Creation & C1 & Integrating information, co-creating content & "[document]Here's the document we've put together" \\
\bottomrule
\end{tabular}
\\[1mm]
\raggedright \footnotesize
\textit{Note.} Interaction levels increase from top to bottom, requiring greater cognitive engagement.
\end{table*}

\subsubsection{The Results of CPS Coding}

Two researchers independently coded all 2,420 utterances. Inter-rater reliability reached $\kappa = 0.830$ ($p < 0.001$), indicating acceptable coding quality. Wayfinding interactions (W1–W3) accounted for 65.29\% of discourse, followed by sense-making interactions (S1–S3, 17.02\%). Creation interactions (C1, 4.46\%) were infrequent. This may be due to the inherent difficulty of generating integrative solutions \citep{huang_research_2020}, or simply because once consensus was reached, teams tended not to spend extended time on further collaborative construction.


\subsection{Synergy Degree Model for CPS} \label{sec:synergy_degree_model}

The SDM is grounded in synergetics, a framework explaining how local interactions produce global organization in complex systems \citep{haken_synergetics_1977}. Although rooted in physics, synergetics has been applied in social and cognitive sciences to explain emergence \citep{gao_empirical_2024, kong_examining_2025}. The SDM centers on three concepts: (1) \textit{order parameters} capture the macro-level coordination state of a system, which in CPS can be interpreted as shared direction or collective focus that cannot be reduced to individual behavior; (2) \textit{subsystems} represent interacting components, which in this study map the CIE framework categories to interaction levels suited to cMOOC collaboration; (3) \textit{synergy degree} quantifies the extent to which group behavior exceeds the sum of individual actions, indicating coherent progress or fragmentation. 

By applying SDM to CPS, we treat different interaction levels—operation ($O$), wayfinding ($W$), sense-making ($S$), and creation ($C$)—as subsystems that contribute to overall collaborative synergy, moving beyond simple behavioral frequency analysis.

For the mean value of each behavioral metric $j$ for each group at each week $t$, standardized calculations were performed based on previous research \citep{diakoulaki_determining_1995}:
$$u(e_{tj}) = \frac{e_{tj} - \beta_{j}}{\alpha_{j} - \beta_{j}}$$
where $\alpha_{j} = \max(e_{j}) \times (1 + 0.05)$ and $\beta_{j} = \min(e_{j}) \times (1 - 0.05)$ represent the upper and lower bounds of the metric $j$ adjusted $\pm 5\%$ across all time periods.

The order parameter for each CPS subsystem $s \in \{O, W, S, C\}$ in a week $t$ is computed as:
$$
u_{s}(t) = \sum_{j \in s} w_j \cdot u(e_{tj})
$$
where weights are determined by $w_j = \frac{p_j}{\sum_{k=1}^{n} p_k}$ and $p_j = \sigma_j \cdot \sum_{k=1}^{n} (1 - r_{jk})$. Here $p_j$ represents the information content of metric $j$, $\sigma_j$ is the standard deviation of metric $j$, and $r_{jk}$ is the correlation coefficient between metrics $j$ and $k$. The weekly order parameter ranges from $[0, 1]$: values closer to $1$ indicate greater internal orderliness within the subsystem.

Based on the weekly order parameter of each CPS subsystem, the weekly CPS synergy degree is defined as:
$$C_t = -\lambda \cdot \sqrt{\left|\prod_{s \in \{O, W, S, C\}} \Delta u_s(t)\right|}, \quad t \geq 2$$
where $\Delta u_s(t) = u_s(t) - u_s(t-1)$ is the change in the order parameter of subsystem $s$ from week $t-1$ to $t$, and the direction parameter $\lambda = 1$ when $\prod_{s \in \{O, W, S, C\}} \Delta u_s(t) > 0$, and $-1$ otherwise. The resulting weekly synergy degree $C_t$ ranges from $[-1, 1]$: negative values signify disorder with asynchronous subsystem development, values near $0$ indicate sub-optimal coordination, and positive values represent synchronised, synergistic development across all interaction levels.
\section{Results}


\subsection{Performance of Automated CPS Behavior Classification Models (RQ1)}

Table~\ref{tab:model_performance_mean_std} reports the performance of nine classification models for automatically identifying CPS behaviors. BERT achieved the best overall results (accuracy $0.507 \pm 0.029$, F1 $0.469 \pm 0.028$), while GPT models showed superior precision ($0.539 \pm 0.004$ for few-shot). Among neural networks, TextCNN performed competitively ($0.484 \pm 0.022$), while classical ML models (SVM, RF) achieved moderate performance. LLMs demonstrated superior semantic understanding for CPS behavior classification.  

Given the complexity of CPS behavior classification with 10 categories, moderate accuracy is expected. GPT’s advantage lies in its higher precision and ability to generate classification rationales, making it well-suited for human–AI collaborative coding. In this approach, AI supports human coders by providing reliable predictions and explanations, while the final coding decisions remain human-driven \citep{natarajan_human---loop_2025}.

\begin{table*}[htbp]
\centering
\small
\caption{Mean and standard deviation of model performance on CPS behavior classification}
\label{tab:model_performance_mean_std}
\begin{tabular}{llcccc}
\toprule
\textbf{Model Type} & \textbf{Model} & \textbf{Accuracy} & \textbf{Weighted F1-score} & \textbf{Weighted Precision} & \textbf{Weighted Recall} \\
\midrule
LLMs & GPT-few-shot    & $0.393 \pm 0.002$ & $0.396 \pm 0.002$ & $\mathbf{0.539 \pm 0.004}$ & $0.393 \pm 0.002$ \\
                  & GPT-zero-shot   & $0.361 \pm 0.002$ & $0.366 \pm 0.003$ & $0.514 \pm 0.005$ & $0.361 \pm 0.002$ \\
                  & BERT            & $\mathbf{0.507 \pm 0.029}$ & $\mathbf{0.469 \pm 0.028}$ & $0.495 \pm 0.023$ & $\mathbf{0.507 \pm 0.029}$ \\
\midrule
Neural Model      & FastText        & $0.301 \pm 0.015$ & $0.189 \pm 0.022$ & $0.240 \pm 0.047$ & $0.301 \pm 0.015$ \\
                  & TextCNN         & $0.484 \pm 0.022$ & $0.452 \pm 0.028$ & $0.486 \pm 0.028$ & $0.484 \pm 0.022$ \\
                  & TextRNN         & $0.379 \pm 0.017$ & $0.306 \pm 0.029$ & $0.277 \pm 0.045$ & $0.379 \pm 0.017$ \\
\midrule
Classical ML      & DT              & $0.418 \pm 0.013$ & $0.381 \pm 0.008$ & $0.401 \pm 0.009$ & $0.418 \pm 0.013$ \\
                  & RF              & $0.447 \pm 0.011$ & $0.411 \pm 0.014$ & $0.435 \pm 0.012$ & $0.447 \pm 0.011$ \\
                  & SVM             & $0.459 \pm 0.016$ & $0.419 \pm 0.020$ & $0.462 \pm 0.014$ & $0.459 \pm 0.016$ \\
                  & KNN             & $0.321 \pm 0.018$ & $0.225 \pm 0.033$ & $0.398 \pm 0.062$ & $0.321 \pm 0.018$ \\
\bottomrule
\end{tabular}
\\[2mm]
\raggedright \footnotesize 
\textit{Note.} GPT refers to \textit{DeepSeek R1}. Values represent mean $\pm$ standard deviation over 5 independent runs (GPT) or 5-fold cross-validation (For others, we used the Python package \textit{pytextclassifier v1.3.9} \citep{xu_pytextclassifier_2022} and the Google Colab T4 GPU). Weighted metrics were used due to class imbalance.
\end{table*}

\subsection{Differences in Order Parameters and Synergy Degree Based on Model- vs. Human-Coded Data (RQ2)}

To examine the potential impact of classification errors on system-level measures, we compared the \textit{order parameters} and \textit{synergy degree} computed from GPT-predicted labels and human-coded labels (Table~\ref{tab:human_gpt_owsc_synergy}). GPT-based results showed a higher S-order parameter ($M=0.370$ vs. $M=0.290$) but a lower C-order parameter ($M=0.190$ vs. $M=0.310$) compared to human coding. Differences in O-order and W-order parameters were minimal. Overall CPS synergy degree was comparable between sources ($SD \approx 0.21$), though weekly patterns varied. 

\begin{figure*}
    \centering
    \includegraphics[width=1\linewidth]{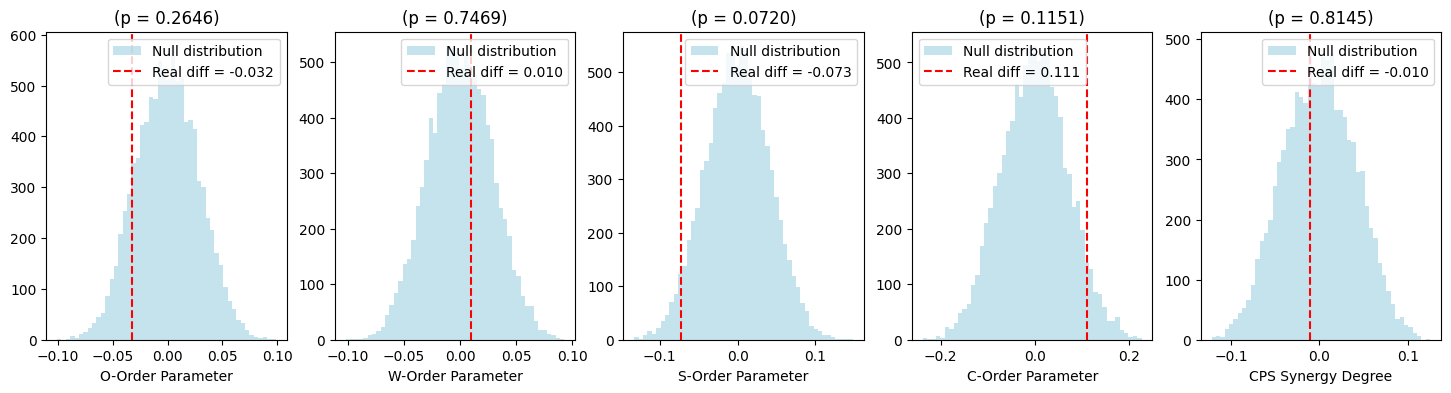}
    \caption{The Permutation Test of CPS Order Parameter and Synergy between Human vs. Model}
    \label{fig:permutation}
\end{figure*}

To statistically assess these differences, we conducted 10,000-iteration permutation tests on the mean differences (Human $-$ GPT; Figure~\ref{fig:permutation}). 
None of the five metrics reached statistical significance ($p > .05$ for all metrics). 
Among them, S-order parameter had the smallest $p$-value ($p=.072$), representing a marginal trend; all others were well within the null distribution. 

Overall, the results suggest that GPT predictions approximate human coding well for system-level metrics but show systematic bias (overestimating sense-making, underestimating creation behaviors). While suitable for exploratory analysis, precise CPS modeling requires human coding, underscoring collaborative AI-in-the-loop approaches.

\begin{table*}[htbp]
\centering
\small
\caption{Descriptive Statistics of Order Parameter and Synergy: Human vs. Model}
\label{tab:human_gpt_owsc_synergy}
\begin{tabular}{rrrrrrrr}
\toprule
\textbf{Week} & \textbf{Group} & \textbf{N}
& \shortstack{\textbf{O-order Param.} \\ (Mean $ \pm $ SD)} 
& \shortstack{\textbf{W-order Param.} \\ (Mean $ \pm $ SD)} 
& \shortstack{\textbf{S-order Param.} \\ (Mean $ \pm $ SD)} 
& \shortstack{\textbf{C-order Param.} \\ (Mean $ \pm $ SD)} 
& \shortstack{\textbf{Synergy Deg.}   \\ (Mean $ \pm $ SD)} \\
\midrule
\textbf{Total}
& GPT   &   55  & $0.293 \pm 0.281$ & $0.442 \pm 0.206$ & $0.370 \pm 0.270$ & $0.190 \pm 0.360$ & $0.036 \pm 0.207$ \\
& Human &   55  & $0.257 \pm 0.284$ & $0.453 \pm 0.161$ & $0.290 \pm 0.309$ & $0.310 \pm 0.383$ & $0.025 \pm 0.211$ \\
\midrule
\textbf{0}
& GPT   &   9  & $0.481 \pm 0.277$ & $0.531 \pm 0.171$ & $0.276 \pm 0.248$ & $0.029 \pm 0.087$ & -- \\
& Human &   9  & $0.412 \pm 0.330$ & $0.442 \pm 0.138$ & $0.120 \pm 0.307$ & $0.001 \pm 0.002$ & -- \\
\midrule
\textbf{1}
& GPT   &   11  & $0.295 \pm 0.205$ & $0.389 \pm 0.226$ & $0.534 \pm 0.260$ & $0.280 \pm 0.437$ & $0.009 \pm 0.073$ \\
& Human &   11  & $0.244 \pm 0.207$ & $0.472 \pm 0.092$ & $0.426 \pm 0.316$ & $0.507 \pm 0.415$ & $0.021 \pm 0.236$ \\
\midrule
\textbf{2}
& GPT   &   11  & $0.288 \pm 0.285$ & $0.458 \pm 0.263$ & $0.455 \pm 0.331$ & $0.240 \pm 0.415$ & $0.102 \pm 0.287$ \\
& Human &   11  & $0.327 \pm 0.329$ & $0.435 \pm 0.192$ & $0.388 \pm 0.354$ & $0.220 \pm 0.370$ & $-0.101 \pm 0.257$ \\
\midrule
\textbf{3}
& GPT   &   12  & $0.340 \pm 0.317$ & $0.359 \pm 0.160$ & $0.284 \pm 0.218$ & $0.193 \pm 0.365$ & $0.085 \pm 0.256$ \\
& Human &   12  & $0.262 \pm 0.273$ & $0.411 \pm 0.209$ & $0.250 \pm 0.234$ & $0.276 \pm 0.349$ & $0.029 \pm 0.161$ \\
\midrule
\textbf{4}
& GPT   &   12  & $0.105 \pm 0.224$ & $0.495 \pm 0.178$ & $0.298 \pm 0.223$ & $0.177 \pm 0.368$ & $-0.022 \pm 0.230$ \\
& Human &   12  & $0.084 \pm 0.218$ & $0.502 \pm 0.149$ & $0.243 \pm 0.292$ & $0.478 \pm 0.399$ & $0.158 \pm 0.214$ \\
\bottomrule
\end{tabular}
\\[2mm]
\raggedright \footnotesize 
\textit{Note.} At week 0, no behaviors were observed in groups G4 and G7, while group G8 exhibited no behaviors in weeks 0, 1, and 2. Accordingly, the sample size was calculated as $N = 5 \times 12 - 5 = 55$.
\end{table*}

\subsection{Groups with different problem types and collaborative quality exhibit significant differences in CPS synergy (RQ3)}

Table~\ref{tab:desc_stats} highlights distinct patterns across problem types and collaborative quality levels. Among problem types, DS groups exhibited the highest O-order parameter ($0.319 \pm 0.248$), suggesting that design-oriented collaboration often requires frequent setup and technical operations in relatively unstructured online learning environments, such as initiating online meetings and sharing screens. MS groups achieved the highest W-order parameter ($0.503 \pm 0.183$), indicating that complex analytical tasks demand intensive interpersonal connections and information sharing. In contrast, SS groups showed the greatest S- and C-order parameters, suggesting that when tasks are relatively simple, learners can more readily devote effort to analytical reasoning and systematic knowledge construction. In terms of collaborative quality, synergy degree displayed a clear stratification: Excellent ($0.084 \pm 0.192$) > Good ($0.059 \pm 0.215$) > Pass ($-0.023 \pm 0.097$) > Fail ($-0.215 \pm 0.316$). 

Overall, the descriptive statistics reveal that groups tackling more difficult tasks (DS, MS) tend to exhibit greater internal orderliness in lower-level CPS subsystems (O and W), whereas groups working on less demanding tasks (SS) are more likely to demonstrate internal orderliness in higher-level CPS subsystems (S and C). Meanwhile, collaborative quality is most strongly reflected in overall synergy degree, underscoring its potential as a sensitive predictor of group learning performance. These preliminary findings suggest that order parameters may be used to characterize the behavioral features and dynamic patterns of group collaboration across different task complexity and structuredness, while synergy degree may serve as an indicator for predicting or explaining the quality of group learning outcomes.

\begin{table*}[htbp]
\centering
\small
\caption{Descriptive statistics of the order parameters and synergy degree across PT and CQ.}
\label{tab:desc_stats}
\begin{tabular}{lrrrrrr}
\toprule
\textbf{Group} & \textbf{N} 
& \shortstack{\textbf{O-order Param.} \\ (Mean $ \pm $ SD)} 
& \shortstack{\textbf{W-order Param.} \\ (Mean $ \pm $ SD)} 
& \shortstack{\textbf{S-order Param.} \\ (Mean $ \pm $ SD)} 
& \shortstack{\textbf{C-order Param.} \\ (Mean $ \pm $ SD)} 
& \shortstack{\textbf{Synergy Deg.}   \\ (Mean $ \pm $ SD)} \\
\hline
\multicolumn{7}{l}{\textbf{PT}} \\
DS & 15 & $\mathbf{0.319 \pm 0.248}$ & $0.401 \pm 0.137$ & $0.204 \pm 0.285$ & $0.262 \pm 0.392$ & $-0.005 \pm 0.168$ \\
MS &  9 &         $0.248 \pm 0.292$  &         $\mathbf{0.503 \pm 0.183}$  & $0.283 \pm 0.322$ & $0.106 \pm 0.317$ & $\mathbf{0.061 \pm 0.182}$ \\
SS & 31 &         $0.230 \pm 0.301$  &         $0.464 \pm 0.163$  & $\mathbf{0.334 \pm 0.317}$ & $\mathbf{0.393 \pm 0.380}$ & $0.029 \pm 0.239$ \\
\hline
\multicolumn{7}{l}{\textbf{CQ}} \\
Excellent & 19 & $0.240 \pm 0.267$          & $0.464 \pm 0.135$          & $0.268 \pm 0.309$          & $0.320 \pm 0.390$           & $\mathbf{0.084 \pm 0.192}$ \\
Good      & 19 & $0.268 \pm 0.293$          & $\mathbf{0.490 \pm 0.164}$ & $\mathbf{0.359 \pm 0.303}$ & $\mathbf{0.393 \pm 0.390}$  & $0.059 \pm 0.215$ \\
Pass      & 12 & $\mathbf{0.269 \pm 0.305}$ & $0.408 \pm 0.161$          & $0.258 \pm 0.287$          & $0.213 \pm 0.357$           & $-0.023 \pm 0.097$ \\
Fail      &  5 & $0.253 \pm 0.348$          & $0.378 \pm 0.231$          & $0.190 \pm 0.426$          & $0.190 \pm 0.426$           & $-0.215 \pm 0.316$ \\
\bottomrule
\end{tabular}
\\[2mm]
\raggedright \footnotesize 
\textit{Note.} At week 0, no behaviors were observed in groups G4 and G7, while group G8 exhibited no behaviors in weeks 0, 1, and 2. Accordingly, the sample size was calculated as $N = 5 \times 12 - 5 = 55$.
\end{table*}

Table~\ref{tab:rq2_sigonly} summarizes the significant results from the ANOVA and Kruskal–Wallis analyses. A main effect of PT was found for the \textit{C-order parameter} ($H=6.45$, $p=.040$), and a main effect of CQ was found for synergy degree ($F=3.44$, $p=.023$). Thus, groups working on different problem types differed significantly in the \textit{C-order parameter}, while groups of varying collaborative quality differed in their overall synergy. These findings support the preliminary patterns observed in the descriptive statistics. Post-hoc comparisons showed that SS groups exhibited higher C-order parameters than MS groups ($U=68.5$, $p=.016$). For collaborative quality, groups with excellent performance showed marginally higher synergy than pass groups ($t=2.052$, $p=.050$).    

Taken together, the results indicate that problem type (e.g., task complexity and structuredness) primarily shapes the C-order parameter, reflecting task-related behavioral structures, whereas collaborative quality is mainly expressed through synergy degree, capturing the overall coordination effectiveness. We will discuss this observation further in Sec. \ref{sec:task-performance-associated}.

\begin{table*}[htbp]
\centering
\small
\caption{Significant group differences in CPS order parameters and synergy degree across PT and CQ factors.}
\label{tab:rq2_sigonly}
\begin{tabular}{lcccr}
\toprule
\textbf{Factor $\times$ Outcome} & \textbf{Test} & \textbf{Statistic (df)} & \textbf{$p$} & \textbf{Post-hoc Comparisons (Mean Diff., $p$)} \\
\midrule
PT $\times$ C-Order Param. & Kruskal--Wallis & $H=6.45$ (2) & $.040^*$ & MS $<$ SS ($-0.287$, $.016^*$) \\
CQ $\times$ Synergy Degree & ANOVA (Fisher)  & $F=3.44$ (3, 51) & $.023^*$ & Excellent $>$ Pass ($+0.107$, $.050^\dagger$) \\
\bottomrule
\end{tabular}
\\[1mm]
\raggedright \footnotesize
\textit{Notes.} 
An asterisk (*) indicates $p < .05$, (†) indicates $p < .1$ (marginal significance). 
The Post-hoc comparisons are performed using the Mann-Whitney U test for non-parametric data, 
and the Welch's t-test for parametric data.
\end{table*}

\section{Discussion}

\subsection{Collaborative AI-in-the-loop CPS Behavior Classification}

Automated coding addresses the limitations of manual analysis, enabling scalable, process-oriented assessment \citep{wang_role_2025}. Our evaluation showed that LLMs outperform other network architectures such as RNNs and CNNs, as well as classical approaches \citep{liu_automated_2022, yan_practical_2024}. While BERT achieved strong performance metrics and may inspire confidence among stakeholders \citep{bansal_does_2021, he_how_2023}, its black-box nature restricts interpretability, providing little explanatory context. In contrast, GPT-based models (e.g., Deepseek R1) generate classification rationales alongside predictions, reducing validation overhead and facilitating human–AI collaboration. As \citet{wang_evaluating_2025} noted, higher accuracy alone does not guarantee trust—limited interpretability may diminish practitioners' willingness to critically examine coding results. GPT’s ability to provide transparent rationales thus encourages human verification and supports a more robust collaborative coding framework.

However, GPT-based models show systematic biases, over-identifying sense-making while under-identifying creation behaviors. This likely reflects the contextual complexity of collaborative creation, where human coders retain an advantage. In connectivist learning environments, creation-level behaviors often manifest through implicit coordination, shared artifacts, or embedded multimedia content that may be less accessible to purely text-based analysis \citep{huang_research_2020}. GPT, relying on surface linguistic cues, struggles with such multimodal signals (this issue may be mitigated through continued improvements in model performance and optimizations of the engineering architecture), whereas human coders leverage holistic knowledge of learners’ processes to provide more accurate judgments. Therefore, these findings highlight the need for collaborative AI-in-the-loop approaches that combine human expertise with AI scalability \citep{natarajan_human---loop_2025, yan_promises_2024}, mitigating the risks of AI autonomy while ensuring reliable interpretation of high-level CPS interactions.

\subsection{Construct Validity of the Automated Synergy Degree Model}

A central concern in educational analytics is whether automated tools faithfully capture the theoretical constructs they are intended to measure \citep{shin_co-coding_2025}. Our validation analysis comparing GPT-predicted and human-coded behavioral sequences addresses this fundamental question for CPS synergy quantification. Permutation tests showed that GPT predictions closely approximated human coding at the system level, with no statistically significant differences across the five examined metrics. Although systematic biases trends were observed at the subsystem level, the preservation of overall synergy degree indicates that the automated approach maintains construct validity for system-level analyses. This finding is important because synergy, as conceptualized in the SDM framework, represents emergent coordination patterns that transcend individual behavioral categories \citep{haken_synergetics_1977, kong_examining_2025}. The convergence of synergy measurements suggests that even when individual behaviors are differentially classified, the underlying coordination dynamics remain reliably captured in automated analysis.

These results support a nuanced interpretation of automated CPS measurement. For exploratory investigations, large-scale screening, or preliminary trend identification, GPT-based synergy quantification appears sufficiently robust. However, when precise characterization of creation-level dynamics or fine-grained synergy patterns is required—particularly in high-stakes assessment contexts—human expertise remains essential. This evidence base reinforces our advocacy for collaborative AI-in-the-loop approaches.

\subsection{Task-Associated and Performance-Associated Patterns in CPS Synergy Emergence} \label{sec:task-performance-associated}
Our other key finding is that the task complexity and structuredness may shape individual order parameters, reflecting task-related behavioral structures, while collaboration quality is mainly expressed through the synergy degree, capturing overall coordination effectiveness. Specifically, in terms of subsystem orderliness, the results showed a significant main effect of problem type on the \textit{C-order parameter}. Post-hoc comparisons further indicated that \textit{survey study} (SS) groups exhibited significantly higher internal orderliness in creation interaction compared to \textit{mode study} (MS) groups. Previous research has indicated that ill-structured problems (e.g., diagnosis, decision-making, design) require learners to integrate multiple perspectives and engage in deeper cognitive processing \citep{jonassen_all_2015}. In cMOOC, generating high-level interactions is key to collective learning \citep{kop_connectivism_2008, wang_interaction_2017}. We found that groups working on the more demanding mode study tasks displayed lower levels of internal orderliness in their creative behaviors. This seemingly counterintuitive task-associated pattern aligns with theoretical perspectives on innovation and creative problem solving, which suggest that a certain degree of “controlled disorder” can foster divergent thinking and exploration, thereby leading to more novel solutions \citep{holling_resilience_1973, lambert_order_2020}. Thus, the relatively low \textit{C-order parameter} observed in MS tasks may not represent negative disorder but rather functional adaptation, reflecting group flexibility in exploring multiple solution pathways and responding to new insights generated through collaborative discovery.

In terms of the emergence of synergy, the relationship between collaborative quality and system-level coordination was primarily reflected in differences in overall \textit{synergy degree} rather than in subsystem \textit{order parameters}. The significant main effect of collaborative quality on synergy degree revealed a clear stratified pattern: \textit{Excellent} > \textit{Good} > \textit{Pass} > \textit{Fail}. This gradient demonstrates a strong correspondence between the level of coordination and collaborative outcomes, with the most pronounced differences observed between high-performing groups and those that struggled. Further, the marginal significance observed in the comparison between \textit{Excellent} and \textit{Pass} groups is particularly noteworthy, as it suggests that synergy degree may serve as a sensitive indicator for distinguishing between successful and borderline collaborative performance. The negative synergy values found in failing groups indicate a lack of coordination among CPS subsystems. This patterns consistent with collaborative breakdown and dysfunctional group dynamics \citep{almaatouq_task_2021}. These performance-associated patterns support the theoretical claim that effective collaboration arises from the coordinated interaction of multiple behavioral subsystems rather than excellence in any single component. This finding resonates with systems theory perspectives \citep{dron_pedagogical_2023, jacobson_conceptualizing_2016, kong_examining_2025}, which emphasize the emergent properties generated through subsystem interactions rather than the additive contributions of individual components.
\section{Conclusion}

This study advanced a novel methodological approach for understanding CPS by integrating automated discourse analysis with system-level synergy modeling. Through NLP techniques and the SDM applied to cMOOC discourse, our investigation yielded three key findings. First, LLMs significantly outperformed other neural network architectures such as RNNs and CNNs, as well as classical approaches for CPS behavior classification. However, GPT showed systematic biases, overestimating sense-making while underestimating creation behaviors. Second, despite classification biases, automated analysis preserved overall synergy measurements, supporting automated quantification for exploratory investigations. Third, \textit{survey study} groups showed higher creation-level ordering than \textit{mode study} groups, suggesting complex tasks require adaptive "disorder," while collaborative quality correlated with synergy degree across performance levels. Our evidence for adaptive value of controlled disorder in complex problem-solving challenges simplistic coordination assumptions and supports complexity-informed collaborative learning design. 

In addition to the methodological contributions, the practical feasibility of applying this approach in authentic learning environments for formative assessment and early intervention. For instructors, atypical parameter patterns may signal the need for timely scaffolding or task redesign. For students, visualizations of synergy trajectories can promote reflection and help groups adjust collaboration strategies.

Several limitations must be acknowledged. First, our group-level focus overlooked personal-level contributions to collaborative synergy. Future investigations should develop multi-level frameworks modeling both personal and group properties. Second, the correlational nature limits causal claims about task characteristics, group processes, and synergy outcomes. Experimental designs manipulating collaborative tasks or support interventions are needed. Finally, the specific context (Chinese cMOOC with WeChat) raises generalizability questions. Replication studies across diverse educational settings, communication technologies, and cultural contexts are needed to establish the broader applicability of the SDM approach.

\begin{acks}
The author gratefully acknowledges the financial support provided by the Society for Learning Analytics Research (SoLAR). 
  This work was also supported by the National Natural Science Foundation of China (NSFC) under Grant Nos. 62577011, 62207003 and 71834002.
The author also gratefully acknowledges the Research Center of Distance Education at Beijing Normal University, as well as the cMOOC development team, for their contributions to course development, cMOOC platform construction, and data collection. 
\end{acks}


\bibliographystyle{ACM-Reference-Format}
\bibliography{content/references}

\section{Appendix}

The prompts used for CPS behavior classification with GPT are as follows\label{app:prompt_example} (The prompt is in Chinese, and the English translation version is in parentheses):

\textit{You are now a text coder in a learning sciences study. Based on the given coding framework, the current message, and the context messages, you need to assign a code to the current message.}

\textit{Coding framework: }
\textit{\{CODEBOOK\}}


\textit{Context: }
\textit{\{CONTEXT MESSAGES\}}

\textit{Current message: }
\textit{\{CURRENT MESSAGE\}}

\textit{Output format: }
\textit{Please output the code only (e.g., W1, S2, C).}

The coding framework is based on the codebook developed in this study. The codebook is embedded in the prompts in LaTeX format and includes four columns: \textit{interaction level, CPS behavior, code, and description}. In the zero-shot task, the \textit{example} column is omitted, whereas in the few-shot task, this column is included. The \textit{CONTEXT MESSAGES} consist of 5 messages that precede the \textit{CURRENT MESSAGE}.

\end{document}